# Value-based Engineering with IEEE 7000[TM]

Sarah Spiekermann & Till Winkler
WU Institute for IS & Society
Vienna University of Economics and Business
Vienna, Austria

*Abstract*—Digital ethics is being discussed worldwide as a necessity to create more reliable IT systems. This discussion, fueled by the fear of uncontrollable artificial intelligence (AI) has moved many institutions and scientists to demand a value-based system engineering. This article presents how organizations can build responsible and ethically founded systems with the "Value-based Engineering" (VBE) approach that was standardized in the IEEE 7000[TM] standard.[1] VBE is a transparent, clearly-structured, step-by-step methodology combining innovation management, risk management, system and software engineering in one process framework. It embeds a robust value ontology and terminology. It has been tested in various case studies. This article introduces readers to the most important steps and contributions of the approach.

## I. INTRODUCTION

In recent years there has been a steady increase in awareness of the need to design technology more ethically. The crashes of two Boeing 737 MAX airplanes as well as the Volkswagen scandal have contributed to the questioning of ethical practices in classic engineering departments. Several studies revealed the potential bias and manipulability of software systems, including critical judicial systems and social networks. The lack of AI transparency became subject to a stream of criticism. To the positive surprise of long-term scholars in the social issues of technology, the IT industry woke up to the need for a more forward-looking, responsible and ethical planning of IT systems.

The progress with approaches that go about more accountable system design, however, have not been without shortcomings. The most important shortcoming is a lack of repeatable, reliable and ethically informed process methodology that allows IT innovation teams (including product managers and engineers) to identify, prioritize and analyze relevant values for their context-specific socio-technical systems and to ensure that these values then find a systematic and verifiable entry into these systems' technical design and development. While such structured methodologies exist to derive classic system and software requirements to get a system up and running (e.g., ISO 15288 or various System Development Life Cycles Models) none exist so far for more value-driven system development.

Of course, there is Value Sensitive Design (VSD), a vary influential approach that has accumulated over 200 case studies, and broadened our knowledge of how values can be explored and conceptualized for IT system design [1]. However, despite its invaluable contribution, VSD has been unable to put forth a stringent process model that companies would be able to follow in a consistent and structured manner. While the iterative coupling of VSD's tripartite "conceptual, empirical and technical" investigations of stakeholder values implies considerable freedom for IT innovation teams, some might desire a clear process and workflow rigor that allows for the traceable translation of transparently derived value principles into concrete system dispositions. As Brent Mittelstadt recently criticized: "the truly difficult part of ethics—actually translating normative theories, concepts and values into 'good' practices … —is kicked down the road like the proverbial can." (p.6 in [2]). What is more, this "translating" of principles into system features seems to benefit from a risk-management logic. Visionary thinkers in the Ethical Computing field have called for a risk-based approach for sustainable system design for a long time [3] and regulators have now started to demand it as well, especially in the face of future AI systems [4].

Value-based Engineering (VBE), the system design approach presented in this article, provides such a risk-based structured methodology to build IT systems with a value focus. Like VSD, VBE develops positive, value-rich visions of what technology could do for society. It embeds positive psychology. Like VSD it also recognizes what can go wrong with technology, what values might be harmed. Unlike VSD, however, it ensures that through a traceable, replicable and iterative process value harms and benefits are systematically and technically addressed and monitored. VBE also recognizes that in order for this to happen, developers' non-value-based functional product roadmaps need to be integrated with value-based requirements. It recognizes that appropriate organizational conditions need to be in place for innovation teams and developers that provide the necessary time, autonomy and ethical guidance. And it embeds a value ontology and value definition that (unlike VSD) is grounded in a post-phenomenological account of value actualization [5, 6].

---

[1] This article solely represents the views of the author and does not necessarily represent a position of either the IEEE P7000 Working Group, IEEE or the IEEE Standards Association. The author of this paper has been vice-chair of IEEE 7000[TM] throughout its entire coming about from 2016-2021. S. Spiekermann, Chair of the Institute for IS & Society, WU Vienna, Welthandelsplatz 1, 1020 Vienna (e-mail: mis-sek@wu.ac.at)

With this contribution VBE also goes beyond what tech co-operations and policy makers currently understand by digital ethics. Over 80 institutions worldwide (including the EU Commission [7] and the OECD [8]) have developed public value commitment lists in the last few years; especially with a view to AI. Indeed, if the five most accepted values on these lists—that is, system transparency, fairness, non-maleficence (safety/security), accountability and privacy—were regularly taught to engineers, fully understood, and rigorously implemented as IT "hygiene factors," then we would certainly witness more people-friendly and social digitization. But these harm-avoiding principle-lists do not allow for saying that a system is really "ethical" in the sense that its purpose is to cater towards "the good." This is what VBE does as it "bases" tech innovators' systems systematically on a positive value mission.

VBE was first sketched out in primitive form in 2016 [9] and then diligently evolved over a five-year standardization process that is called the IEEE 7000$^{TM}$ standard, "A Model process for addressing Ethical Concerns during System Design" [10, 11]. The diligence of the development process has been presented in this magazine previously [11]. In the following I will broadly describe how VBE with IEEE 7000$^{TM}$ works, what its most important constructs are, what literature it is based on, how it relates to other influential work on values and ethics in system design and what its own challenges are.

## II. PREREQUISITES FOR VALUE ELICITATION

To grasp what can be understood by ethical IT design, consider the following example of how voice assistants can operate completely differently depending on the culture in which they are developed: In 2017, when a user said "I am so sad," a U.S. Alexa device replied "I wish I had arms to cuddle you." The Russian counterpart Yandex, on the other hand, replied to the same statement "No one said life is a fun event" [12]. Reading these two completely different answers raises the question: Have the developers of the dialogue systems actually consciously thought about the ethical implications of the systems' answers? Have they considered that these express distinct values and that––depending on the diffusion of the system—these could significantly influence users' attitude to life, for instance, children growing up with the system? In fact: Which answer would actually be the more correct one? The American or the Russian one? Thinking about this last question goes to the heart of ethical system design: "How should I act?" [13] "I," that is, the engineer of the dialogue system. How do I give the voice assistant the "right," "good" or "wise" dialogue? The answer to this question is a function of the values one wants to pass on with a system.

In this example, it is up for discussion whether the voice assistant should rather promote the virtue of mental toughness, personality robustness and discipline—or focus on feeling good, conveying closeness and compassion. The decision taken has an important value-ethical impact on society once the respective voice assistant is used at massive scale. It should therefore be answered by innovation teams with a great sense of responsibility; a responsibility that takes time, ethical guidance and work conditions often not provided these days to system developers [14].

VBE intends to provide for this. If organizations sign up for it and seek compliance with IEEE 7000$^{TM}$, then processes are established to officially grant engineering and innovation teams the time to engage in value reflection. IEEE 7000$^{TM}$ outlines in its introduction that the following organizational conditions must also be met (p.9 in [10]):

— Readiness to include a wide group of stakeholders in the engineering effort
— An open, transparent and inclusive project culture
— A commitment to quality
— A dedication to ethical values from the top of the organization
— A commitment to allocate sufficient time and resources for ethical requirements definition

Guided by a new professional called a "Value Lead" a fine-grained value structure of the future SOI is developed with stakeholders. Value Leads "contribute subject matter expertise and facilitative skills, bridging the gap between engineering, management and ethical values in a constructive way" (p. 33 in [10]). They are not "the person in charge of ethics" (p. 33 in [10]), because VBE foresees that all innovation team members, product managers, system developers and management co-operate to design a system. But Value Leads know how to run organizations through this new form of system development life cycle. They should have substantial ethical knowledge, know the conceptual and technical details of widely accepted value principles (like privacy, transparency, etc.), understand Material Value Ethics and moral philosophy well enough to apply them in VBE projects and make them fruitful for system development, of which they should have a good grasp.

Value Leads must ensure that organizations are not primed by any existing value-principle lists. The voice assistant example shows that the context and nature of this system implies the recognition of many values not included in any of the globally accepted value-commitment-lists (e.g., closeness, compassion, personal robustness, discipline…). This discrepancy between that which is ethically essential for a "system of interest" (SOI) in a context and that which is globally and generically listed as essential is significant. Moreover, value-lists can "prime" value elicitation. A group of students at WU Vienna analyzed an indoor location-tracking system for a fashion retailer targeting senior customers. One student group identified privacy as the most important ethical challenge of this tracking system. Prior to their analysis they had read a value-principle list including privacy. In contrast, a control group of students that did not read the list but physically visited the fashion store to talk to the elderly customer stakeholders learned something else: They discovered that the most essential core value the location tracking system could offer is "helpfulness." Helpfulness for the customers in getting service on the spot. They also learned that the value of privacy becomes secondary for seniors at the moment where these get a value like helpfulness back in exchange for sharing their location data.

The core value of "helpfulness" could be actualized in this case through various value qualities, such as quick and convenient access to sales associates, improved orientation in the store, or time savings. The first phase of VBE is all about understanding this physical deployment context and what stakeholders might value in a technology. Note that in describing the value sphere VBE benefits from a value ontology and very precise vocabulary for value elicitation that becomes apparent in this example. A core value is a high intrinsic value "that is identified as central in the context of a SOI" (p. 17 [6, 10]) whereas a value quality (also called value "demonstrator" in IEEE 7000$^{TM}$) is "a potential manifestation of a core value, which is either instrumental to the core value or undermines it" (p. 23). A core value (like helpfulness) can actualize through a value quality (like convenient service access) when there is a respective value disposition built into a system. A value disposition, such as a mobile retailer app with a big green "help button" for fashion customers to summon staff (and that uses the location data) is a "system characteristic that is an enabler or inhibitor for one or more values" (p. 23) at the level of the digital thing.

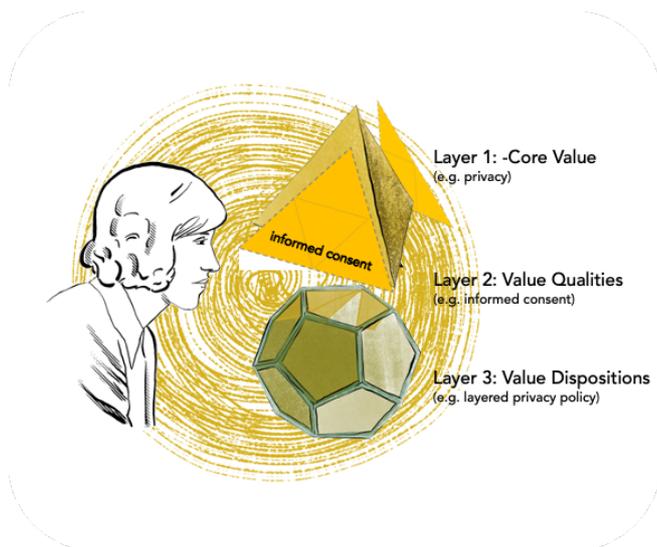

Fig. 1. VBE's three-layered value ontology (taken from [6])

This three-layered value ontology underlying VBE and IEEE 7000$^{TM}$ is depicted in Figure 1. It is derived from Max Scheler's opus magnum, "Formalism in Ethics and Non-formal Ethics of Values" [5]. It considers everything surrounding us—other people, nature, technology, relationships, or activities—as potential carriers of value. Things like a voice assistant or a location tracking application can become bearers of value if they embed technical configurations that enable these. The configuration of the voice assistant dialogue, for example, is the prerequisite for either the value of compassion to be revealed to a user or the value of toughness. For the senior in the fashion store, the value of "helpfulness" may unfold when she looks at the location app on her smartphone and sees a big green button that allows staff to be summoned. Thinking about "value bearers," "value dispositions," "core values" and "value qualities" allows us to think about a system design from the beginning in a value-strategic and thereby ethical way.

### III. MATERIAL VALUE ETHICS AND MORAL PHILOSOPHY ARE THE ROOT OF VALUE-BASED ENGINEERING

Going back to the location-tracking system two critical questions arise: First, how does VBE ensure that a value strategy is really ethical? Especially when, for the reasons mentioned above, one does not initially resort to predefined, institutionally approved lists of values or established norms. And second: Does the value elicitation phase of IEEE 7000$^{TM}$ de-prioritize such a recognized value as privacy behind a rather practical value such as customer "helpfulness"? The following sections will answer these questions

#### A. Use of Moral Philosophies for the Exploration of Values

When eliciting values with IEEE 7000$^{TM}$ three recognized moral theories are used. With reference to the respective SOI and with an initial Concept of Operation in mind, three questions are asked, which are derived from the moral philosophies of utilitarianism [15], virtue ethics [16], and duty ethics [13]:

- utilitarianism: what human, social, economic, or other values are affected, positively or negatively, by the SOI if that system were used at scale?

- virtue ethics: what is the long-term impact of the SOI on the personal character of the affected stakeholders if that system were used at scale?

- duty ethics: what personal maxims or value priorities does the project team see affected by the SOI that the project team members believe are so important that they wish to preserve them in society?

These three questions are elaborated not only by the project team directly responsible for building the SOI, but by a broad group of critical stakeholder representatives. The discourse between stakeholders as well as their selection should meet requirements such as those envisioned by Jürgen Habermas in his Discourse Ethics [17]. Furthermore, if a culture in which the SOI is to be used has a specific ethical orientation embedded in it that goes beyond these three ethical theories, then the IEEE 7000$^{TM}$ standard encourages a fourth question for that culture, which is grounded in its respective philosophical or spiritual tradition.

The result of such a morally guided reflection on values is that even for simple SOIs, a relatively broad spectrum of values is identified. Across three case studies, we empirically observed that each stakeholder identified an average of 16–19 values [18]. For a Viennese telemedicine start-up, innovation management students identified a total of 54 unique value violations that could arise from the platform, as well as 63 positive values promoted [18]. This large spectrum of values, both positive and negative, makes one aware of how ethically fine-grained and sensitive technology cases are when scrutinized in this way.

To cope with this complexity, VBE structures the values it finds. Core values are those that are repeatedly described and seem to carry special stakeholder weight. They are typically intrinsic, which means that they are desired for their own sake, leaving little doubt as to what they are good for [19]. Complementary are the value qualities instrumental to them. These also result first from the stakeholder dialogues. For example, if a stakeholder is concerned about the privacy of his voice assistant, he might say that he does not want the voice assistant's speech protocols to be sold, or that there should be no unauthorized recording, that data security must be guaranteed, etc. Such stakeholder statements demonstrate the contextual "qualities" of the core value (privacy), which are seen bottom-up and should therefore be respected in the SOI's dispositions (Figure 2, left).

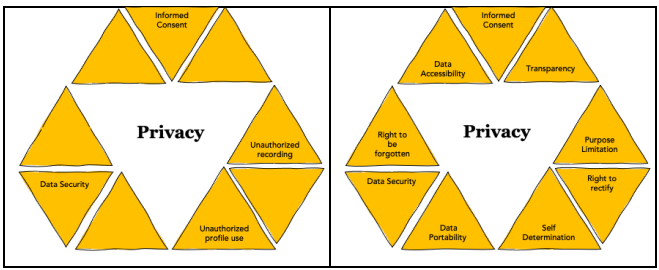

Fig. 2. Comparison of core privacy cluster derived from stakeholders (left) and in completed form after conceptual analysis (right)

The value qualities collected in this way are usually not complete. Stakeholders have intuitive access to values through their "value feelings." But these are not enough to fully conceptualize a core value. The right-hand image in Figure 2 shows this. Anyone familiar with the privacy principles considered in the EU's General Data Protection Regulation knows that there are significantly more. For this reason VBE engages is what Value Sensitive Design has coined "conceptual analysis" [20]. In a separate analysis, VBE completes the core value qualities in line with the law and the philosophical literature, often respecting the wording that is chosen in those more established sources (Figure 2, right).

*B. Prioritizing Core Value Clusters*

When an organization comes up with 10 (or sometimes more) core value clusters, which ones should it prioritize? VBE with IEEE 7000™ does not envision that core values will be pitted against each other. They will not be treated as trade-offs. Instead, the clusters are presented to executives and stakeholder representatives to rank them in terms of their importance to the organization. This ranking, which will later determine the priorities in system development, considers the following criteria: (p. 41 in [10]):

1. "Stakeholders agree that the SOI is good for society and avoids unnecessary harm.
2. The organization does not use people merely as a means to some end.
3. Organizational leaders can accept responsibility for the value priorities chosen according to their own personal maxims.
4. The organization respects its own stated ethical organizational principles if there are any.
5. The organization can commit to the value priorities in its business mission.
6. The environment is maximally preserved.
7. The organization considers existing ethical guidelines."

These prioritization criteria embed both Kant's duty ethics [13] as well as organizational and external value expectations. External value expectations are, for example, values enshrined in laws, industry commitments or international agreements (such as the value principles described above), the United Nations Convention on Human Rights or the EU Charter of Fundamental Rights. So if a stakeholder team has ranked the value of customer "helpfulness" (in the fashion house example) before that of customer privacy, then aligning with the legal and political externals at this point ensures that privacy might be ranked before helpfulness, or at least ranked high enough to ensure the system's compliance with such external value expectations from the start.

IV. FROM PRINCIPLES TO PRACTICE

If a company decides to invest in an SOI, the next question is how the value strategy can be systematically incorporated into the system design. To this end, each contextual "core value (CV)/value quality (VQ)" tuple is translated into a so-called Ethical Value Requirement (EVR). An EVR is defined in IEEE 7000™ (p.18) as an "organizational or technical requirement catering to values that stakeholders and conceptual value analysis identified as relevant for the SOI." Thus, the EVR is in its nature more concrete than a value quality. It is the "bridge" between the values in the world that are relevant to the SOI and the concrete, specific system-level requirements (SR) that will guide the enabling of these in the SOI (Figure 3).

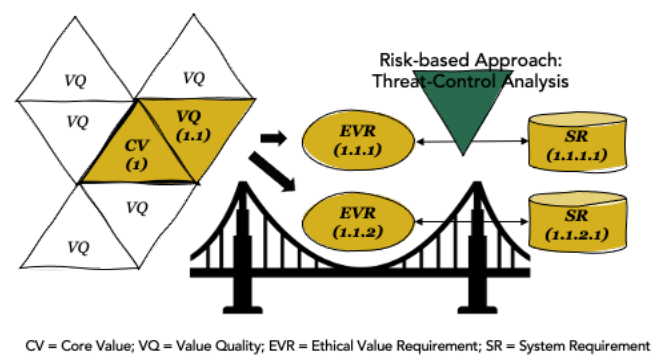

CV = Core Value; VQ = Value Quality; EVR = Ethical Value Requirement; SR = System Requirement

Fig. 3. EVRs are the bridge connecting the ethical value world to the system engineering world (taken from [6])

As an example, take the core value of privacy, a value quality of which might be "informed consent." One EVR could read: "Ensure that a user can give consent to his/her data processing in an easy and informed way, whereby 'informed' means that the information provided is instantly accessible and comprehensible for laypeople." When EVRs are put down in what IEEE 7000™ calls a Value Register, each of them should be concretized and qualified with adjectives such as "easy," "instantly accessible," "comprehensible for laypeople" or even give concrete, quantified and testable thresholds, assumptions or constraints. Such thresholds can later also be used for certification and testing of the system. Does the system meet the EVR thresholds set out by the innovation team?

EVRs are the starting point for deriving system-level requirements. However, the above example shows that EVRs are not only of technical but can also be of organizational nature. The comprehensibility with which an informed consent is programmed requires a prior organizational decision on the openness and honesty of customer communication. Some EVR measures even need to be taken independently of any technical features. For example, one indispensable measure to ensure the helpfulness of a retailer's indoor tracking app besides the technology itself is to hire sufficient staff to deal with customer help-calls resulting from the app's use. Such purely organizational EVR measures are the reason for path I in Figure 4 (bottom left). This first immediate organizational management path makes plain that VBE is not only a system developer challenge, but one where the administrative leadership of an organization must work hand-in-hand with the technical units.

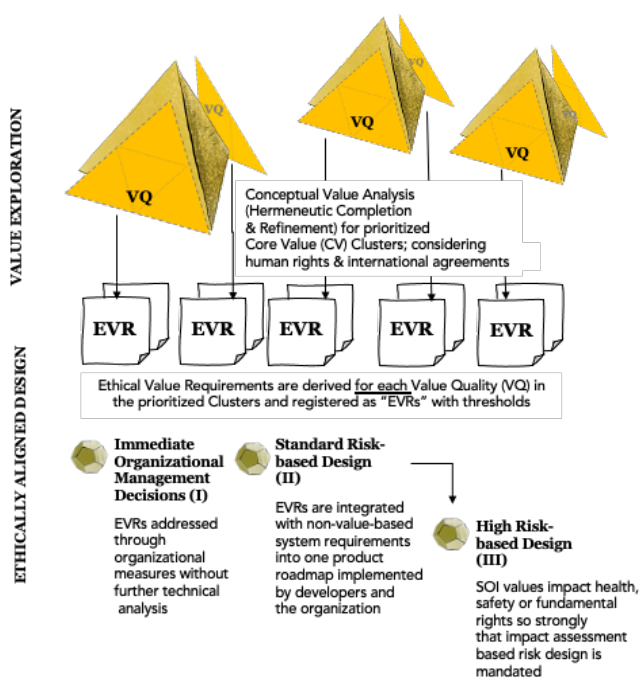

Fig. 4. Rough flow of constructs for ethical system requirements derivation

Where organizational management action does not suffice, VBE offers two alternative risk-based technical processes for system design. This is shown at the bottom of Figure 4: What I call a "Standard Risk-based Design Process" (in line with IEEE 7000™'s section 9) and an "Impact assessment-led System Design Process," as many high-risk systems require one (and has been standardized elsewhere, such as [21]). The former applies to probably most cases in system design.

The latter comes into play when particularly critical values are at stake (such as health or security), when particularly risky systems are built or when the regulator mandates an impact assessment-led system design.

What a standard risk-based design approach does is that it asks how each EVR could be put at risk of not being fulfilled. The risk of nonfulfillment comes from concrete "threats." Technical system requirements, so-called "controls," are then developed to address each threat in such a way that the EVR can be fulfilled. In this way potential value threats are technically mitigated before they happen.

When a high-risk design path (III) is necessary, then this analysis is further deepened. The impact or damage potential is analyzed for all those system values that need special care. The probability of threats is scrutinized. Depending on the damage potential and the likelihood of a threat, corresponding controls are chosen and residual risk is documented (for more information see [6]).

The overall reason for both risk-based approaches is that by conceiving of values as "being-at-risk," development teams are put into a spirit of care and awareness long called for by computer ethicists.

As Figure 3 illustrates, the chain from core values, to value qualities, to EVRs and their threat and controls (system requirements) are traced through a numbered chain documented in the Value Register. In this way organizations are able to show their compliance with prioritized values. They can recap why they took or failed to take certain value-based decisions. When system requirements are entered into the product roadmap and implemented, this chain can be continued. As a result, an organization keeps track of its ethical maturity. It remembers how value-based system features were prioritized in comparison to non-value-based ones. And it also records whether the system's validation period confirmed value creation and/or effective value breach prohibition. When value qualities don't actualize as planned or unexpected ones appear, this can be monitored in the Value Register and another EVR analysis with subsequent risk-based design can be triggered.

V. CONLUSION AND CHALLENGES FOR VALUE-BASED ENGINEERING

The description of VBE with IEEE7000™ shows that ethical engineering is not an isolated corporate process that can be outsourced to Compliance Management, Risk Management or the CSR Department. Instead, it is an integrated, strategic exercise that drives the technical "value proposition" at the heart of an organization. For this reason it is recommended that the earliest possible phases of product design are accompanied by VBE. One case study with a telemedicine company [22] showed that if VBE is implemented too late, when an SOI's business purpose is already fixed, then management's readiness to change the value mission in line with VBE's ethical exploration is limited.

This does not mean that VBE is only suited to early start-up phases. It can also be applied to brownfield situations. But then the organization employing it must be prepared to heavily invest. This is because VBE typically leads—our case studies show [22, 23]—to a very different technology narrative, often requiring an adapted architecture.

Another challenge is that a VBE organization must involve its key service partners. The ecosystem around an SOI—the so-called "system-of-systems"—must be part of the VBE analysis to avoid unexpected and uncontrollable undermining of values. This again means that SOI operators may have to forgo some preferred supply chain partners. IEEE 7000™ requires partners who are ready to give access to their systems, which is often not the case in today's highly distributed service architectures.

Sometimes there will not be reliable enough partners and an organization needs to build a technology block by itself. The product roadmap that developers work on is hence heavily influenced. In VBE organizations the roadmap is driven just as much by value-based priorities as by non-value-based requirements in bringing a system up and running. This is reflected in the debriefing interview of one of our VBE case studies with UNICEF's Yoma platform, where the CTO noted that VBE would bring an "80–85% change in terms of how you design a system" [23]. Still, he added, he would "definitely use it again," especially because he felt that VBE would, crucially, foster stakeholder value rather than the normal "founder-mentality."

That said, investors need to be ready to give their money for the stakeholder "good" and not every capital provider is likely to embrace this. Three case studies additionally showed, however, that service creativity is significantly fueled through VBE [18]. Many more product ideas and value potentials are unveiled. And—not surprisingly—there is 10 times more value harm detection than in today's ordinary development approaches [18]. So, while VBE requires a fundamental rethinking of the way we approach systems development, it also leads to more sustained investments.

## Acknowledgment

I want to thank the IEEE 7000™ CRG group and especially Lewis Gray and Ruth Lewis for all their ideas and recommendations. I also want to acknowledge the tireless support of my Ph.D. students Kathrin Bednar and Till Winkler who accompanied me in this research endeavor.

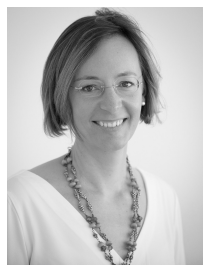
S. Spiekermann chairs the Institute for IS & Society at Vienna University of Economics and Business since 2009. She has authored over a hundred scientific articles, including publications in IEEE Proceedings, IEEE Transactions on Software Engineering, IEEE Security and Privacy, Communications of the

ACM, Journal of Information Technology (JIT), and many other leading outlets. She is author of the textbook "Ethical IT Innovation – A Value-based System Design Approach" as well as the books Networks of Control (2016) on personal data markets and Digital Ethics – A Value System for the 21$^{st}$ Century (2019). Sarah is a member of IEEE and has been co-initiator and vice-chair of IEEE 7000$^{TM}$ from 2016–2021.

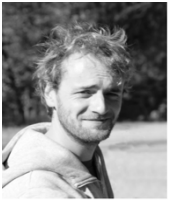

Till Winkler holds a master and bachelor in psychology and is a Ph.D. student at the Institute for IS & Society at Vienna University of Economics and Business. He was an IEEE 7000 workgroup member from 2016-2021